\documentstyle[prl,aps,epsfig]{revtex}
\begin{document}
\input{psfig.sty}
\def \gam {\frac{ N_f N_cg^2_{\pi q\bar q}}{8\pi} }
\def \gamm {N_f N_cg^2_{\pi q\bar q}/(8\pi) }
\def \be {\begin{equation}}
\def \ba {\begin{eqnarray}}
\def \ee {\end{equation}}
\def \ea {\end{eqnarray}}
\def \gap {{\rm gap}}
\def \gapp {{\rm \overline{gap}}}
\def \gappp {{\rm \overline{\overline{gap}}}}
\def \im {{\rm Im}}
\def \re {{\rm Re}}
\def \Tr {{\rm Tr}}
\def \P {$0^{-+}$}
\def \S {$0^{++}$}
\def \uu {$u\bar u$}
\def \dd {$d\bar d$}
\def \ss {$s\bar s$}
\def \qq {$q\bar q$}
\def \qqq {$qqq$}
\def \si {$\sigma(500-600)$}
\def \lsm {L$\sigma $M}
\title{A note on the baryonic  $B\to\bar{\Lambda} p\eta^\prime$ decay}
\author{F. Piccinini\footnote{On leave of absence from INFN, 
Sezione di Pavia, Pavia, Italy} and A.D. Polosa
}
\address{Theory Division, CERN, CH-1211, Geneva 23, Switzerland}
\maketitle
\begin{abstract}
In this short note we examine the exclusive three-body
$B\to\bar{\Lambda}p\eta^\prime $ decay using a simple pole 
model involving a scalar intermediate resonance state. 
Our aim is to test the recently formulated hypothesis 
that charmless baryonic $B$ decays could occur mainly 
in association with $\eta^\prime$ or $\gamma$.
\\
\vskip 0.05cm \noindent Pacs numbers: 13.25.Hw, 12.39.-x,
14.40.Cs\\ \noindent CERN-TH/2001-379; FNT/T 2001/24; hep-ph/0112294\\
\end{abstract}

In a recent paper by Hou and Soni~\cite{Hou}, the general problem
of searching new ways to estimate charmless baryonic $B$ decays
is addressed. The thesis is that 
charmless baryonic $B$ decays may be more prominent in association 
with $\eta^\prime$ or $\gamma$. In particular, the attention is 
posed on the exclusive
process $B\to \eta^\prime \bar{\Lambda} p$, taking the cue from the
experimental observation of the unexpectedly large modes $B\to
\eta^\prime X_s $ and $B\to\eta^\prime K$~\cite{exp}. 
 
Since the enhancement
($Br(B\to \eta^\prime K)\simeq 8\times 10^{-5}$)  was established 
by the CLEO collaboration,
many studies aimed at investigating its nature have appeared.
An interesting proposal to explain the phenomenon is that based on the
subprocess
$b\to sg^*\to s\eta^\prime g$, where the virtual gluon $g^*$
emerging from the standard model penguin couples to $\eta^\prime$
via an effective $g g^* \eta^\prime$ vertex related to the gluonic
triangle anomaly~\cite{atwood}. The structure of this vertex was
reexamined in ref.~\cite{hou}, where the running of the effective
coupling of $\eta^\prime$ to gluons, assumed to be constant in
ref.~\cite{atwood}, is also taken into account. The possibility that
the $g g^* \eta^\prime$ vertex could be dangerously affected by
out of control nonperturbative effects was  discussed in
ref.~\cite{atwood2}. Some further criticism can be found in
ref.~\cite{petrov}. The emergence of a $g^* g\eta^\prime$ 
coupling in the $B\to D\eta^\prime$ decay has 
been explored in ref.~\cite{miod}.
In ref.~\cite{ahmadi} a `non-spectator model'  involving 
a gluon fusion process has been introduced to study the inclusive 
$B\to \eta^\prime X_s$ and the exclusive $B\to K^{(*)}\eta^\prime$ 
decays: the  gluon $g$ of the $g g^*\eta^\prime$ vertex is supposed 
to be emitted by the light quark inside the $B$ meson, while 
the $g^*$ comes from the $b\to s$ penguin.

Taking advantage from the latter mechanism, we estimate the $B\to
\eta^\prime \bar{\Lambda} p$ branching ratio using a simple pole model
according to which this decay proceeds via an intermediate {\it
scalar} meson. A pole model is used also in ref.~\cite{Hou} to gain a
quantitative estimate of $B\to \eta^\prime \bar{\Lambda} p$, but
the intermediate  state there assumed is a $K$ meson which makes
the pole approximation questionable because
the $K$ is clearly quite off its mass shell. As it's already noted in 
\cite{Hou} it would be preferable to exploit the idea of a $g^*$ emerging 
from the penguin and fragmenting into a diquark pair rather than
rely on the simple picture of an intermediate state mediating the 
baryonic decay. The former approach takes care of the short distance 
dynamics which is instead completely lost when considering only the long
distance contribution due to the intermediate state. Diquark models 
and sum rules are certainly the most 
complete approaches to baryonic decays (see the discussion 
in \cite{ultimo}, see also \cite{book}), anyway, in many cases, 
simple pole ideas have provided reasonable estimates of 
several exclusive processes.

In this note we  will consider a pole model of the 
$B\to \eta^\prime \bar{\Lambda} p$ interaction involving 
as intermediate meson state the $K_0^*(1430)$ scalar resonance~\cite{pdg}; 
in other words we will assume that the decay proceeds as follows
$B\to \eta^\prime K_0^* \to \eta^\prime \bar{\Lambda} p $. The 
effective couplings $K\bar{\Lambda} p$ and $K_0^* \bar{\Lambda} p$ have been
computed in ref.~\cite{rijken} in the framework of a Nuclear-Soft-Core
model. It is interesting to observe that the latter coupling
is suggested to be almost ten times bigger than the former
(see Tables VI and VII in ref.~\cite{rijken}), suggesting that
the $K_0^*(1430)$ state is a quite better candidate for being considered
as the intermediate state in $B\to \eta^\prime \bar{\Lambda} p$.

In Fig. 1 it is shown the diagram we are considering while in Fig. 2
we report the gluon fusion mechanism supposed to be
responsible for the $\eta^\prime$ coupling to the $B$ meson \cite{ahmadi}.
The penguin interaction and the
$\eta^\prime g g $ vertices are depicted effectively
as two  black spots, while the interaction of the almost-on-shell
gluon (carrying momentum $p$) with the light quark line is
represented with a smaller spot. This `non-spectator-mechanism' has been
used in ref.~\cite{ahmadi} to predict the $Br(B\to K\eta^\prime)$ branching
ratio. In this note we merely use the model to fit $B\to K\eta^\prime$
and afterwards to predict $B\to K_0^*\eta^\prime$. Once the
$B K_0^*\eta^\prime$ coupling is known, using the pole model we
estimate  the $Br(B\to \eta^\prime \bar{\Lambda} p)$ with the Breit-Wigner
approximation for the intermediate $K_0^*$.

The effective $\eta^\prime gg $ vertex is given by:
\begin{equation}
A^{\mu\sigma}(gg\to \eta^\prime)=i H(q^2,p^2,m_{\eta^\prime}^2)
\epsilon^{\mu\sigma\alpha\beta}q_\alpha p_\beta,
\end{equation}
where the form factor $H(0,0,m_{\eta^\prime}^2)$ is estimated to
be approximately $1.8$ GeV$^{-1}$~\cite{atwood}. Here we
consider the $q^2$ dependence of $H$ described in ref.~\cite{ali} (see
Fig. 13 in ref.~\cite{ali}), where $q^2=m_{\eta^\prime}^2+2p_0
E_{\eta^\prime}$ (see Fig. 2). Our starting point is the
expression for the amplitude of $B\to K \eta^\prime$ obtained in
ref.~\cite{ahmadi}:
\begin{eqnarray}
\langle \eta^\prime K|H_{\rm eff}|B\rangle &=&-i\frac{2 C H f_B
f_K}{9\Lambda^2}(p_B\cdot q\; p_K\cdot p-p_B\cdot p\; p_K\cdot
q)\nonumber\\ &=&-i\frac{2 C H f_B f_K}{9\Lambda^2} m_B
p_0\left[(m_B-E_K)E_K-\frac{(m_B^2-m_\eta^2-m_K^2)}{2}\right],
\label{eq:prima}
\end{eqnarray}
where:
\begin{equation}
H_{\rm eff}=i C H (\bar{s}\gamma_\mu(1-\gamma_5)T^a
b)(\bar{q}\gamma_\sigma T^aq) \frac{1}{p^2}
\epsilon^{\mu\sigma\alpha\beta}q_\alpha p_\beta.
\end{equation}
The $C$ constant is built with the Inami-Lim function $E$~\cite{inami} 
according to:
\begin{equation}
C=\frac{G_F}{\sqrt{2}}\frac{\alpha_s}{2 \pi}
V_{tb}V_{td}^*[E(x_t)-E(x_c)],
\end{equation}
where:
\begin{eqnarray}
E(x_i)&=&-\frac{2}{3}\ln(x_i)+\frac{x_i^2(15-16x_i+4x_i^2)}{6(1-x_i)^4}\ln(x_i)\nonumber\\
&+&\frac{x_i(18-11x_i-x_i^2)}{12(1-x_i)^3},
\end{eqnarray}
$x_i=m_i^2/m_W^2$, $m_i$ being the internal quark mass and we assume
$\alpha_s=0.2$, $f_B=0.2$ GeV, $f_K=0.167$ GeV.

The second
equation in eq.~(\ref{eq:prima}) is obtained in the center of mass frame of the
decaying $B$ and averaging on the directions of the gluon radiated
by the light quark in the $B$ system (see Fig. 2). Obviously:
\begin{eqnarray}
E_K&=&\sqrt{m_K^2+{\bf p}_K^2}\nonumber\\
|{\bf p}_K|&=&\left[ \frac{m_B^2+m_K^2-m_{\eta^\prime}^2}{4 m_B^2}-
m_K^2\right]^{\frac{1}{2}}. \nonumber\\
\end{eqnarray}

Assuming the cutoff $\Lambda =0.23$ GeV in (\ref{eq:prima}) 
($\Lambda \approx \Lambda_{\rm QCD}$), and considering also $p_0$,
the energy of the almost-on-shell gluon emitted softly by
the light quark, $p_0\approx \Lambda$, we reproduce fairly well
the experimentally observed central value  of $Br(B\to
K\eta^\prime)=7.8\times 10^{-5}$.

Having fitted the  parameter $\Lambda$ (see Fig. 3) 
on the observed rate for $B\to K\eta^\prime$, once $p_0$ is chosen to be
$p_0 \simeq 0.25$ GeV, we can now consider
the case of the $B\to K_0^*\eta^\prime$ process writing, after
ref.~\cite{ahmadi}, the expression for its amplitude as:
\begin{equation}
\langle \eta^\prime K_0^* | H_{\rm eff} | B\rangle =+ i\frac{2 C H
f_B f_{K_0^*}}{9\Lambda^2}(p_B\cdot q\; p_{K_0^*}\cdot p-p_B\cdot p
\; p_{K_0^*}\cdot q).\label{eq:seconda}
\end{equation}
We take the definition and the value of the leptonic decay
constant $f_{K_0^*}$ in ref.~\cite{maltman}:
\begin{equation}
f_{K_0^*}m_{K_0^*}^2=0.0842\pm 0.0045 \;\;\; {\rm GeV^3}.
\end{equation}

Using the amplitude given in eq.~(\ref{eq:seconda}) we readily  compute
the branching ratio:
\begin{equation}
Br(B\to K_0^*\eta^\prime)=3.4\times 10^{-6}.
\end{equation}
To perform this estimate we use a value for the $H$ form factor
given by $H=1.5$ GeV$^{-1}$ instead of the
$H(0,0,m_{\eta^\prime}^2)=1.8$ GeV$^{-1}$. This is because we take
into account the form factor suppression extensively described in
ref.~\cite{ali} (see Fig. 13 of ref.~\cite{ali}).

The latter prediction is  functional to compute the $Br$ for the
process $B\to\bar{\Lambda}p\eta^\prime$ using the coupling
\begin{equation}
\frac{g_{K_0^*\Lambda p}^{\rm (eff)}}{\sqrt{4\pi}}=-2.83,
\end{equation}
computed in ref.~\cite{rijken}. (In ref.~\cite{rijken} it is used 
the symbol $\kappa$ rather than $K_0^*$; see 
note~\cite{rijkennote}). The
expression for the width is given by:
\begin{eqnarray}
&&\Gamma(B\to\bar{\Lambda}p\eta^\prime)_{K_0^*}=\nonumber\\&& 
\frac{G_1 G_2}{8(2\pi)^3 m_B^3}\times
\int_{(m_\Lambda+m_p)^2}^{(m_B-m_{\eta^\prime})^2}dq^2
\lambda^{\frac{1}{2}}(m_B^2,q^2,m_{\eta^\prime}^2)
\frac{1}{(m_{K_0^*}^2-q^2)^2+\Gamma_{K_0^*}^2
m_{K_0^*}^2}
\frac{1}{q^2}\lambda^{\frac{1}{2}}(q^2,m_\Lambda^2,m_p^2),
\label{eq:ultima}
\end{eqnarray}
where $G_2$ and $G_1$ are respectively $(g_{\kappa\Lambda
p}^{\rm (eff)})^2$ and $|\langle \eta^\prime K_0^* | H_{\rm eff} |
B\rangle|^2$. The couplings $BK_0^*\eta^\prime$ and 
$B(K_0^*)_{\rm off-shell}\eta^\prime$ are assumed to be the same  and 
the Breit-Wigner formula is implemented.
Eq. (\ref{eq:ultima}) allows for a prediction of the
branching ratio:
\begin{equation}
Br(B\to\bar{\Lambda}p\eta^\prime)=1.2\times 10^{-6},
\label{quella}
\end{equation}
which is sensibly lower than what expected in ref.~\cite{Hou}, namely 
$Br(B\to\bar{\Lambda}p\eta^\prime)=2.4\times 10^{-5}$.
Varying smoothly the value of $p_0$ and  selecting accordingly the
value of the form factor $H$ from ref.~\cite{ali} and the value of the
cutoff $\Lambda$ in order to fit $B\to K\eta^\prime$,
we find a nice stability of the $Br$ obtained in eq.~(\ref{quella}).
Even more stable against variation of the parameters is the value of
$Br(B\to\eta^\prime K_0^*)$.
It is worth noting that
$\Gamma(B\to\bar{\Lambda}p\eta^\prime)_{K_0^*}/\Gamma(B\to K_0^*\eta^\prime)
\simeq 0.36$.
\noindent
However, it should be stressed that, due to the intrinsic model dependence 
of our  approach and due to the complexity of the baryonic decays, 
these results have to be understood  as  order-of-magnitude estimates.
Interestingly, what emerges here is that a simple pole model, suggests 
a quite reasonable rate for a charmless baryon-antibaryon final state
produced in association with $\eta^\prime$ in a $B$ decay.

This exclusive mode could be soon reconstructed by the BaBar
and Belle  collaborations and our straightforward calculation provides the
possibility to test the non-spectator model in ref.~\cite{ahmadi}, the
Nuclear-Soft-Core model in ref.~\cite{rijken} and, more basically, the
anomaly picture in ref.~\cite{atwood}.

Once observed, $B$ meson baryonic modes could also offer new 
paths to explore fundamental topics such as the extraction 
of  CP violating phases.

\acknowledgements We would like to thank M.L. Mangano, 
Th.A. Rijken and D.O. Riska for useful discussions. ADP acknowledges 
support  from the Marie Curie fellowship programme, 
contract HPMF-CT-2001-01178.


\begin{figure}[t!]
\begin{center}
\epsfig{bbllx=0.5cm,bblly=16cm,bburx=20cm,bbury=23cm,height=4.8truecm,
        figure=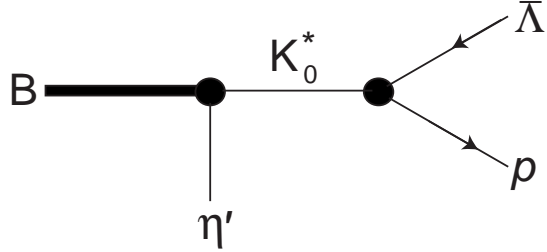}
\caption{\label{fig} \footnotesize The
$B\to\bar{\Lambda}p\eta^\prime$ decay is modeled to proceed via
the intermediate scalar resonance $K_0^*(1430)$. With respect to
the calculation sketched in ref.~[1], where the $K$ is taken in place
of $K_0^*$, here we are considering the more reliable case in
which the intermediate  state is not heavily off its mass
shell. Moreover the effective coupling of $K_0^*\bar{\Lambda}p$,
calculated in ref.~[9],
is definitely stronger than that of $K\bar{\Lambda} p$.}
\end{center}
\end{figure}


\begin{figure}[t!]
\begin{center}
\epsfig{bbllx=0.5cm,bblly=16cm,bburx=20cm,bbury=23cm,height=4.8truecm,
        figure=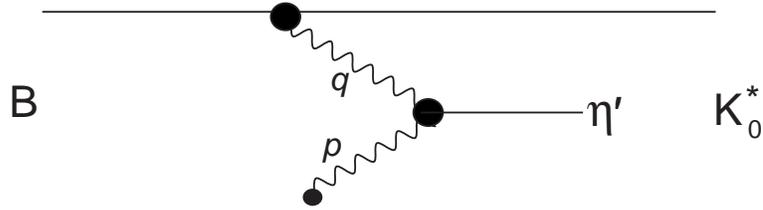}
\caption{\label{fig} \footnotesize According to the
non-spectator-model the $\eta^\prime$ is produced via the gluon
fusion of the virtual gluon produced in the SM $b\to s$ penguin
and the soft gluon radiated by the light quark in the $B$. }
\end{center}
\end{figure}


\begin{figure}[t!]
\begin{center}
\epsfig{
height=5.0truecm,
        figure=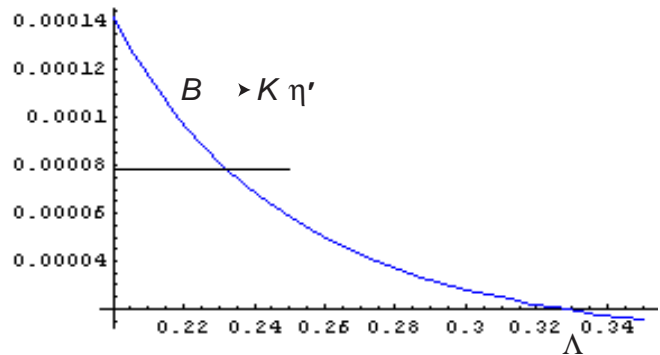}
\caption{\label{fig} \footnotesize In the non spectator model
there is an important dependence on the cutoff $\Lambda$ chosen.
In our case we use $\Lambda$ just as a fit parameter. We fit the
$B\to K\eta^\prime$ to predict $B\to K_0^*\eta^\prime$. The
experimental value indicated by the horizontal line selects 
$\Lambda=0.23$ GeV. }
\end{center}
\end{figure}


\begin{thebibliography}{10}
\bibitem{rijkennote} The $\kappa$ state has actually the same
quantum numbers as $K_0^*(1430)$ except for the mass which is
taken to be $1300$ MeV in the paper by Stocks and Rijken~\cite{rijken}. 
One can take into account this mass difference by
considering a $10\%$ variation in the value of the effective
coupling.
\end{thebibliography}
\end{document}